Author for correspondence:
J. Salmon
e-mail: julien@boulder.swri.edu

# Accretion of the Moon from non-canonical discs

J. Salmon and R. M. Canup

Department of Space Studies, Southwest Research Institute, 1050 Walnut Street, Suite 300, Boulder, CO 80302, USA

Impacts that leave the Earth–Moon system with a large excess in angular momentum have recently been advocated as a means of generating a protolunar disc with a composition that is nearly identical to that of the Earth's mantle. We here investigate the accretion of the Moon from discs generated by such 'non-canonical' impacts, which are typically more compact than discs produced by canonical impacts and have a higher fraction of their mass initially located inside the Roche limit. Our model predicts a similar overall accretional history for both canonical and non-canonical discs, with the Moon forming in three consecutive steps over hundreds of years. However, we find that, to yield a lunar-mass Moon, the more compact non-canonical discs must initially be more massive than implied by prior estimates, and only a few of the discs produced by impact simulations to date appear to meet this condition. Non-canonical impacts require that capture of the Moon into the evection resonance with the Sun reduced the Earth–Moon angular momentum by a factor of 2 or more. We find that the Moon's semi-major axis at the end of its accretion is approximately $7R_\oplus$, which is comparable to the location of the evection resonance for a post-impact Earth with a 2.5 h rotation period in the absence of a disc. Thus, the dynamics of the Moon's assembly may directly affect its ability to be captured into the resonance.

## 1. Introduction

The origin of the Earth's Moon remains an unsolved problem. The most favoured scenario involves a giant impact on the forming Earth that placed material into orbit from which the Moon accreted [1,2].

The *canonical* case involves the oblique impact of a Mars-size object at velocities of order $10\,\mathrm{km\,s^{-1}}$ [3–10], resulting in the formation of an approximately 1.5–2$M_L$ disc ($M_L$ is the mass of the Moon), with more than 20% of its material exterior to the Roche limit, $a_R \approx 2.9 R_\oplus$, and a vapour fraction of order 10% (e.g. [9,10]). $N$-body simulations of a particulate protolunar disc showed that the Moon would accrete on a time scale of order 1 year, and with an average semi-major axis approximately $1.3 a_R$ [11,12]. Such fast accretion would imply an initially fully molten Moon, potentially at odds with geological constraints [13] and GRAIL measurements [14], and would not allow time for equilibration of the protolunar disc with the Earth's atmosphere, which has been proposed as a possible explanation for the striking isotopic similarities between the Earth and Moon [15]. However, the initial protolunar disc is not a purely condensate disc and is instead expected to be a two-phase mixture of silicate melt and vapour.

A self-gravitating disc can fragment into clumps due to instabilities if it has a sufficiently high surface density and the dispersion velocity between the disc components is sufficiently low. A magma protolunar disc would probably be gravitationally unstable and fragment into clumps. Exterior to the Roche limit, clumps can be stable and accumulate to form the Moon. Interior to the Roche limit, clumps are constantly sheared apart by planetary tides, resulting in an increased rate of collisions that generate a substantial viscosity and heat the disc [16]. This heating is sufficient to largely vaporize the disc, rendering it gravitationally stable [17,18]. As a vapour disc radiatively cools from its surfaces, magma begins to condense and gravitational instability is re-instated, either due to the reduction in sound speed associated with a two-phase mixture if the disc vapour and magma remain vertically well mixed or due to instability in a magma mid-plane layer if the disc vertically stratifies. Thus, instability in a condensed disc heats and vaporizes the disc, but as the disc vaporizes it stabilizes, which allows the vapour disc to cool and re-condense. This feedback mechanism suggests that the viscosity of the disc is regulated by the rate at which it radiates energy, resulting in an evolution over approximately $10^2$ years [17,18].

Salmon & Canup [19] studied the accretion of the Moon from discs similar to those produced by canonical impacts using a hybrid model, in which material located inside the Roche limit is represented by a simplified fluid disc that viscously evolves with a time scale appropriate for a two-phase silicate protolunar disc [17], while material outside the Roche limit is tracked with a standard $N$-body accretion model. They found that the Moon forms in three consecutive steps: (i) material beyond $a_R$ rapidly forms a large, sub-lunar-sized moonlet that confines the inner disc inside $a_R$ via resonant interactions, (ii) the inner disc viscously spreads outwards and reaches the Roche limit after several tens of years, and (iii) new moonlets are spawned by inner disc material spreading beyond the Roche limit, and these are accreted by the Moon over a period of approximately $10^2$ years. This more protracted accretion time scale could allow sufficient time for the composition of the inner protolunar disc to equilibrate with that of the Earth [15]. Both scattering of spawned moonlets by the Moon and trapping of inner moonlets into mean motion resonances with the Moon generally lead to a net positive torque on the Moon's orbit, causing the Moon to form at a typical orbital distance of $2 a_R$ [19], substantially larger than found in pure $N$-body simulations.

Recently, new kinds of Moon-forming impacts have been explored that involve (i) a very large impactor containing about half of the Earth's mass [20] or (ii) a higher velocity, sub-Mars-sized impactor that collides nearly head-on into a rapidly spinning Earth [21]. These *non-canonical* impacts produce a planet–disc system with 2–2.5 times the angular momentum in the current Earth and Moon. Subsequent prolonged capture of the Moon into the evection resonance with the Sun could drain enough angular momentum from the Earth–Moon system to make it compatible with its current value [21]. Capture into evection requires that the Moon is initially interior to the resonance and crosses it at a sufficiently slow rate [21,22]. In addition, a relatively narrow range of tidal parameters in the Earth and Moon may be required for the resonance to remove enough angular momentum to make the non-canonical impacts suitable lunar formation candidates [21].

These high-angular-momentum impacts produce more compact (i.e. most of the mass is concentrated close to the Earth) discs than canonical impacts. In canonical impacts, gravitational torques across the distorted shape of the oblique impactor immediately after the initial impact are highly effective at placing some of the distant impactor material into extended orbits [9]. The emplacement mechanism in the high-angular-momentum cases involves the ejection of mass effectively from the surface of the Earth, which is rotating at near the fission rate (either due to a prior impact in the Ćuk & Stewart [21] model or due to the Moon-forming impact in the Canup [20] model), a mechanism that is less effective at torquing material into high orbits. A key finding of [19] is that incorporation into the Moon of disc material located inside the Roche limit is rather poor, so that it is unclear whether these more compact discs will be successful at forming lunar-mass objects.

We here apply the model developed in [19] to study the accretion of the Moon from non-canonical discs. In §2, we present the numerical model. In §3, we present results of numerical simulations of non-canonical discs. Implications are discussed in §4.

## 2. The model

### (a) Roche-interior disc

The numerical model we use is identical to that in [19]. Inside the Roche limit, the disc is described as a uniform surface density 'slab' that extends from the Earth's surface out to an outer edge $r_d$, where we typically assume $r_d = a_R = 2.9 R_\oplus$ for the initial disc. We assume for simplicity that the liquid and vapour phases coevolve. This is probably a good assumption for the disc model of Thompson & Stevenson [17], and a poor assumption for the stratified disc model of Ward [18]. Initially, the disc spreads with a viscosity $\nu_{TS}$ set by the limiting rate at which a silicate vapour photosphere can radiate away the viscously generated heat [17],

$$\nu_{TS} = \frac{\sigma_{SB} T_p^4}{\sigma \Omega^2}, \tag{2.1}$$

where $\sigma_{SB} = 5.67 \times 10^{-8}$ W m$^{-2}$ K$^{-4}$ is the Stefan–Boltzmann constant, $T_p$ is the temperature of the disc's photosphere, $\sigma$ is the disc's surface density (taken to be that of both the vapour and liquid components), $\Omega = \sqrt{GM_\oplus/r^3}$ is the Keplerian frequency at distance $r$ and $G$ is the gravitational constant. The viscous spreading of the disc is performed by computing the rate of change of its edges using the viscosity evaluated at the disc's outer edge, $r_d$.

As the disc spreads and loses mass both onto the planet and outwards through the Roche limit (see below), the viscosity estimated by the above expression becomes larger than that physically expected for a fully molten disc subject to local gravitational instabilities [16]. At this point, the rate of viscous heat generation probably falls below the rate at which the vapour disc can radiatively cool, and the silicate vapour component of the disc can condense. The then fully magma disc will evolve with a viscosity $\nu_{WC}$ set by transport of angular momentum by transient gravitational instabilities in the magma that continuously form and are then disrupted by planetary tides [16]

$$\nu_{WC} = \frac{\pi^2 G^2 \sigma^2}{\Omega^3}. \tag{2.2}$$

As material spreads outwards past the Roche limit via viscous spreading, it can accumulate into small objects that are no longer destroyed by planetary tides. When the inner disc extends beyond $a_R$, we compute the mass $m_f$ of the fragment that would form from gravitational instabilities [23]

$$m_f = \frac{16\pi^4 \xi^2 \sigma^3 r_d^6}{M_\oplus^2}, \tag{2.3}$$

where $\xi$ is a coefficient of order unity that we set to 0.3, and $r_d$ is the position of the inner disc's outer edge. If the disc's outer edge is at the Roche limit ($r_d = 2.9 R_\oplus$), then inner discs containing 1.5 and 0.01$M_L$ will form fragments of $\approx 3 \times 10^{-3} M_L$ and $10^{-9} M_L$, respectively. The new body is then added to the N-body part of the code (see below) and the disc's mass and outer edge are adjusted so as to conserve mass and angular momentum.

## (b) Roche-exterior disc

Beyond the Roche limit, gravitational instabilities in a magma disc may lead to the rapid formation of clumps that are stable against tidal disruption. This means that, in this region, instabilities may not produce a large and ongoing source of viscous dissipation, as they do in the Roche-interior disc (see appendix A). In this case, although the material in the outer disc may initially be a mixture of melt and vapour, it will probably rapidly cool and condense (see also Discussion). We thus model the material beyond the Roche limit as a collection of individual particles. Orbital evolution of the objects is performed using the $N$-body symplectic integrator Symba [24], which we have modified to include interactions with the inner disc. The latter is done by computing, for each orbiting object, the 0-th order (i.e. 2:1, 3:2, etc.) Lindblad resonances that fall into the disc. The associated torque $T_s$ is [19,25]

$$\frac{T_s}{M_s} = \frac{\pi^2}{3} \frac{M_s}{M_\oplus} G \sigma a_s \sum_{m=2}^{m^*} 2.55 m^2 \left(1 - \frac{1}{m}\right), \qquad (2.4)$$

where $M_s$ and $a_s$ are the mass and semi-major axis of the orbiting body, and $m^*$ is the highest $m$ for which resonance $(m:m-1)$ falls into the disc. When a body orbits close to the disc, we only consider the resonances that fall exterior to the body's Hill radius. The acceleration $\mathbf{a}_{res}$ of the body due to the interaction with the disc is computed using the formalism of Papaloizou & Larwood [26], $\mathbf{a}_{res} = \mathbf{v}/t_m$, where $\mathbf{v}$ is the body's velocity, and $t_m = L_s/T_s$ is the body's orbital migration time scale and $L_s$ its angular momentum.

To determine whether collisions between orbiting particles should result in a merger, we use tidal accretion criteria [27,28], with normal and tangential coefficients of restitution set to 0.01 and 1. Finally, if a body passes within $2R_\oplus$, we assume it is tidally disrupted, following the pericentre criterion of Sridhar & Tremaine [29]. When this happens, we remove the object from the $N$-body code, and add its mass and angular momentum to that of the inner disc.

## 3. Accretion from non-canonical discs

### (a) Simulation protocol

Table 1 shows the initial parameters of our simulations, chosen based on successful cases in [20,21], defined as post-impact discs having compositions that deviate from that of the silicate Earth by less than or equal to 15% (assuming a 'Mars-like' composition impactor), together with disc masses large enough to produce a lunar-mass Moon based on prior accretion efficiency estimates of Ida *et al.* [11]. The successful cases in the 'fast spinning Earth' model of Ćuk & Stewart [21] have an average disc mass $\langle M_d \rangle = 2.3 M_L \pm 0.4$, and an average normalized disc-specific angular momentum, defined as $J_d \equiv L_d/(M_d \sqrt{GM_\oplus a_R})$ where $L_d$ is the total disc angular momentum, of $\langle J_d \rangle = 0.85 \pm 0.05$. Successful cases in the 'half-Earth impactor' model of Canup [20] have $\langle M_d \rangle = 3.1 \pm 1.3$ and $\langle J_d \rangle = 0.90 \pm 0.06$. Thus, the discs in the half-Earth impact are on average more massive and somewhat more radially extended than those in the fast spinning Earth model.

We set the initial number of particles in the outer disc to 1500. The initial surface density profile of the outer disc is $\sigma(r) \propto r^{-q}$, where we have set $q = 5$, motivated by results from [20]. We consider initial disc parameters that span the range of outcomes observed in table 1 of [20,21]. We select only the discs for which the analytical criterion of [11] predicts the formation of a moon with a mass more than or equal to $0.95 M_L$. Results from the two papers overlap, but with the latter producing discs with masses generally smaller than the former, so that runs 1–21 are mostly representative of results from [21], while runs 7–27 are mostly representative of results from [20]. We use total initial disc masses ranging from 1.75 to $3.25 M_L$, with inner disc masses of 1.5, 2 and $2.5 M_L$ and outer disc masses of 0.25, 0.5 and $0.75 M_L$. Because our model inner disc assumes a uniform surface density with distance from the planet, we can only adjust its specific angular momentum by shifting its outer edge. We do this in runs 28–31 to reach the lowest values of specific angular momentum considered here. Overall, we consider discs with a total specific angular momentum ranging from 0.857 to 0.925 (in units of $\sqrt{a_R GM_\oplus}$). This reflects a lower average value (because the discs are more compact) than what we considered for canonical discs, where we used discs with an initial total specific angular momentum ranging from 0.843 to 1.099 (in units of $\sqrt{a_R G M_\oplus}$) [19].

**Table 1.** Simulation parameters. $M_d$ and $L_d$ are the disc's initial total mass and angular momentum, respectively. $M_{in}$ and $M_{out}$ are the disc's mass inside and outside the Roche limit, respectively. The Roche-interior disc initially extends from $1R_\oplus$ to $a_{in}^{max}$. The initial outer disc extends from $a_{in}^{max}$ to $a_{out}^{max}$. Units of mass, distance and angular momentum are the present lunar mass $M_L$, Earth radius $R_\oplus$ and angular momentum of the Earth–Moon system $L_{EM} = 3.5 \times 10^{34}$ kg m$^{-2}$ s$^{-1}$.

| run | $L_d/M_d$ ($\sqrt{a_R GM_\oplus}$) | $L_d$ ($L_{EM}$) | $M_d$ ($M_L$) | $M_{in}$ ($M_L$) | $M_{out}$ ($M_L$) | $a_{in}^{max}$ ($R_\oplus$) | $a_{out}^{max}$ ($R_\oplus$) |
|---|---|---|---|---|---|---|---|
| 1 | 0.871 | 0.275 | 1.75 | 1.5 | 0.25 | 2.9 | 3.5 |
| 2 | 0.875 | 0.277 | 1.75 | 1.5 | 0.25 | 2.9 | 4 |
| 3 | 0.878 | 0.277 | 1.75 | 1.5 | 0.25 | 2.9 | 4.5 |
| 4 | 0.892 | 0.322 | 2 | 1.5 | 0.5 | 2.9 | 3.5 |
| 5 | 0.899 | 0.325 | 2 | 1.5 | 0.5 | 2.9 | 4 |
| 6 | 0.904 | 0.327 | 2 | 1.5 | 0.5 | 2.9 | 4.5 |
| 7 | 0.909 | 0.369 | 2.25 | 1.5 | 0.75 | 2.9 | 3.5 |
| 8 | 0.918 | 0.373 | 2.25 | 1.5 | 0.75 | 2.9 | 4 |
| 9 | 0.925 | 0.376 | 2.25 | 1.5 | 0.75 | 2.9 | 4.5 |
| 10 | 0.865 | 0.351 | 2.25 | 2 | 0.25 | 2.9 | 3.5 |
| 11 | 0.868 | 0.353 | 2.25 | 2 | 0.25 | 2.9 | 4 |
| 12 | 0.870 | 0.354 | 2.25 | 2 | 0.25 | 2.9 | 4.5 |
| 13 | 0.882 | 0.398 | 2.5 | 2 | 0.5 | 2.9 | 3.5 |
| 14 | 0.888 | 0.401 | 2.5 | 2 | 0.5 | 2.9 | 4 |
| 15 | 0.892 | 0.403 | 2.5 | 2 | 0.5 | 2.9 | 4.5 |
| 16 | 0.897 | 0.445 | 2.75 | 2 | 0.75 | 2.9 | 3.5 |
| 17 | 0.904 | 0.449 | 2.75 | 2 | 0.75 | 2.9 | 4 |
| 18 | 0.910 | 0.452 | 2.75 | 2 | 0.75 | 2.9 | 4.5 |
| 19 | 0.861 | 0.428 | 2.75 | 2.5 | 0.25 | 2.9 | 3.5 |
| 20 | 0.863 | 0.429 | 2.75 | 2.5 | 0.25 | 2.9 | 4 |
| 21 | 0.865 | 0.430 | 2.75 | 2.5 | 0.25 | 2.9 | 4.5 |
| 22 | 0.876 | 0.475 | 3 | 2.5 | 0.5 | 2.9 | 3.5 |
| 23 | 0.880 | 0.477 | 3 | 2.5 | 0.5 | 2.9 | 4 |
| 24 | 0.884 | 0.479 | 3 | 2.5 | 0.5 | 2.9 | 4.5 |
| 25 | 0.889 | 0.522 | 3.25 | 2.5 | 0.75 | 2.9 | 3.5 |
| 26 | 0.895 | 0.525 | 3.25 | 2.5 | 0.75 | 2.9 | 4 |
| 27 | 0.900 | 0.528 | 3.25 | 2.5 | 0.75 | 2.9 | 4.5 |
| 28 | 0.857 | 0.348 | 2.25 | 2 | 0.25 | 2.2 | 4 |
| 29 | 0.860 | 0.350 | 2.25 | 2 | 0.25 | 2.4 | 4 |
| 30 | 0.854 | 0.424 | 2.75 | 2.5 | 0.25 | 2.2 | 4 |
| 31 | 0.857 | 0.426 | 2.75 | 2.5 | 0.25 | 2.4 | 4 |

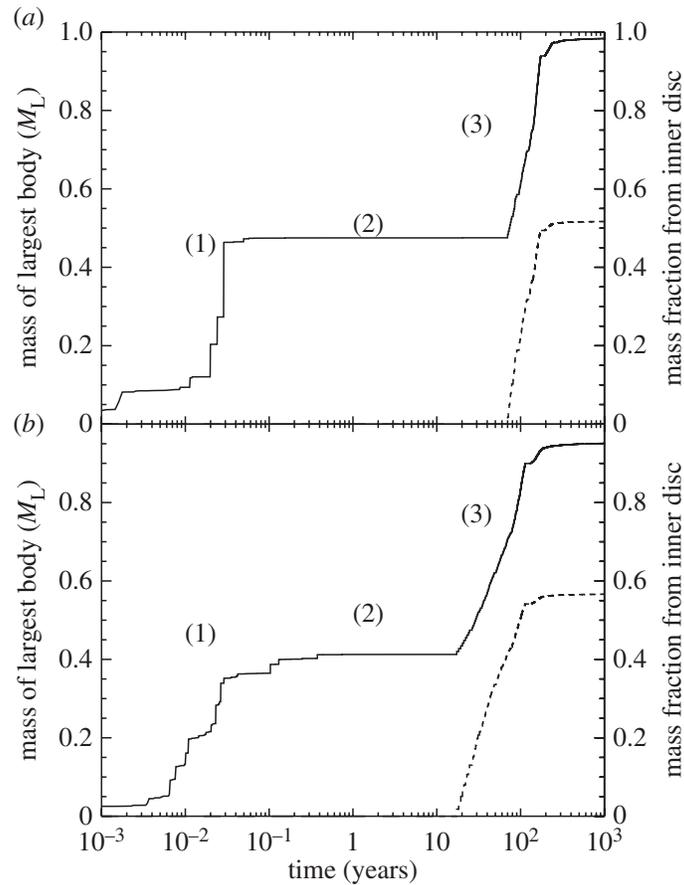

**Figure 1.** Evolution of the mass of the largest body (solid line) and the fraction of its mass derived from Roche-interior disc material (dashed line), for run 22 (*a*) and for a sample canonical case (run 34 in [19], *b*). Material from the Roche-interior disc is accreted in a protracted phase after several tens of years, due to its being confined early on by bodies in the outer disc.

## (b) Results

### (i) Accretion dynamics

Figure 1*a* shows the evolution of the mass of the largest body in run 22, and the fraction of its mass that consists of material accreted from the Roche-interior disc. For comparison, figure 1*b* shows the same quantities for a sample canonical disc whose initial outer disc contains the same total mass ($0.5 M_L$), but whose inner disc is initially less massive ($2 M_L$ versus $2.5 M_L$; see fig. 3 in [19]). Figure 2*a* shows the number of orbiting bodies in run 22, with again figure 2*b* showing the same quantity for the canonical case (fig. 4 in [19]). Figure 3 shows the mass and position of the outer edge of the Roche-interior disc.

For both cases, accretion occurs in three phases. In phase 1, bodies in the outer disc collide and accrete, forming one massive body on a time scale of order approximately 1 year (figure 1, solid line). The inner disc is confined inside the Roche limit due to resonant interactions with outer bodies (figure 3, solid line). In phase 2, the inner disc slowly viscously spreads outwards. After approximately 70 years, the inner disc reaches the Roche limit, and in phase 3 new bodies are spawned and continue the accretion of the Moon. One difference is that, for the non-canonical case, phase 3 starts somewhat later (approx. 70 years) than in the canonical case (approx. 20 years), due to both the higher inner disc mass in the non-canonical case (which decreases $\nu_{TS}$ and thus increases the disc's spreading time scale) and strong confinement of the inner disc in the non-canonical case (the outer disc being more compact, the objects within it orbit closer to the inner disc and thus exert a stronger resonant torque onto it).

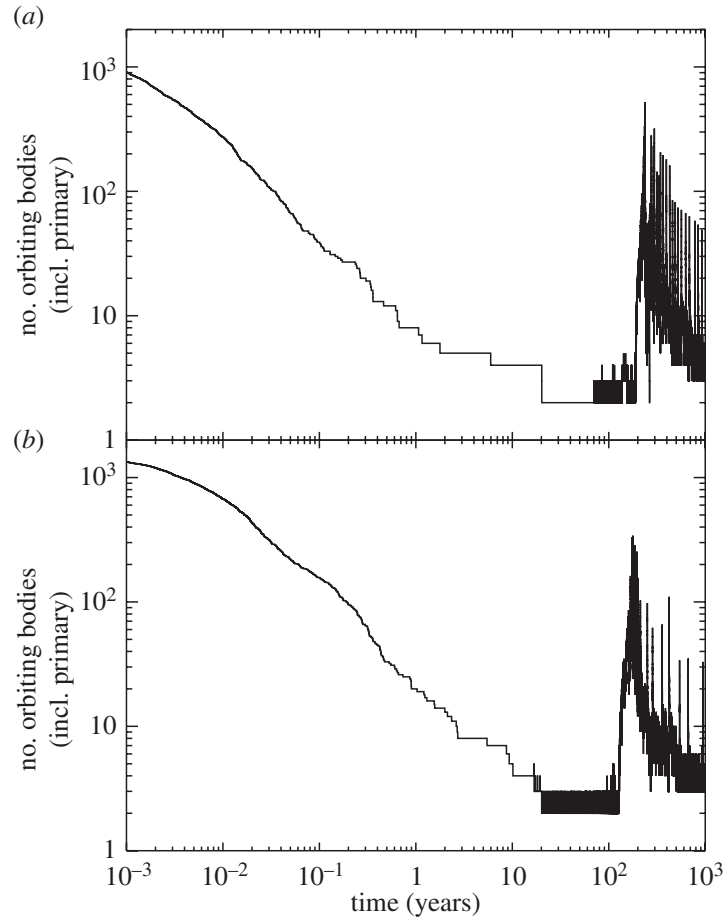

**Figure 2.** Evolution of the number of orbiting bodies in run 22 (*a*) and for a sample canonical disc (run 34 in [19], *b*). The number of orbiting bodies decreases initially as objects in the outer disc collide and merge or scatter one another. After several tens of years, the Roche-interior disc has spread back out to the Roche limit and starts producing new moonlets.

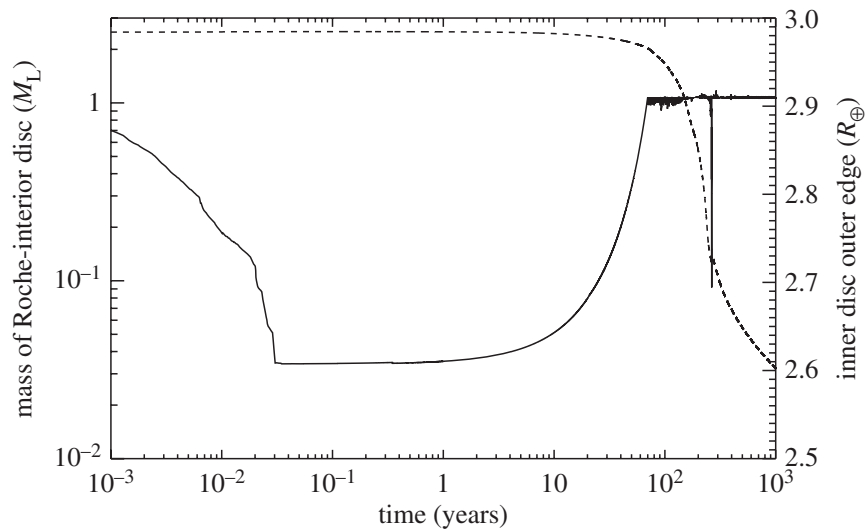

**Figure 3.** Evolution of the mass of the Roche-interior disc (dashed line) and the position of its outer edge (solid line), in run 22. Resonant interactions with outer bodies initially confine the disc inside the Roche limit $a_R = 2.9 R_\oplus$. At $t \sim 70$ years, the disc reaches $a_R$ and starts forming new moonlets that repeatedly confine the disc inside $a_R$. At $t \sim 250$ years, a moonlet is scattered inwards by the Moon, causing the disc's outer edge to move inwards until the confining body gets tidally disrupted (once its pericentre gets within $2R_\oplus$).

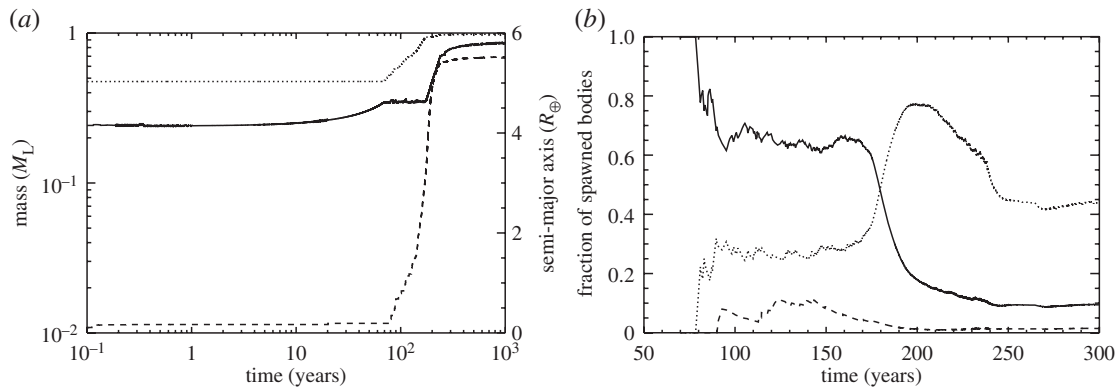

**Figure 4.** (a) Mass and semi-major axis of the Moon (dotted and solid lines, respectively), and cumulative mass of bodies tidally disrupted and absorbed into the inner disc after being scattered towards the Earth (dashed line). The Moon initially sits at approximately $4.2R_\oplus$, at which point it still has Lindblad resonances into the inner disc. As the latter viscously spreads, the Moon's semi-major axis expands until the inner disc reaches $a_R$. At approximately 170 years, the resonant torque from the inner disc is too weak to excite the moonlets' semi-major axes faster than the Moon is exciting their eccentricities. As a result, they get scattered towards the Earth and tidally disrupted. The Moon gains their angular momentum and its semi-major axis expands. (b) Fraction of bodies spawned from the inner disc that either merge with the Moon (solid line), get scattered inwards and tidally disrupted (dotted line) or are ejected from the system (dashed line). As the disc mass decreases, the rate of increase of the moonlets' semi-major axes decreases while their eccentricities are increasingly excited by the growing Moon, resulting in an increasing number of objects being scattered inwards and a net expansion of the Moon's semi-major axis.

In both cases, there is an accretion 'stall' around $t \sim 150$ years, during which the Moon's semi-major axis expands while its mass remains relatively constant (figure 4a, solid line). When the Moon accretes a moonlet soon after it is spawned at the Roche limit, the Moon's semi-major axis decreases because the moonlet has a lower specific orbital angular momentum than the Moon itself. On the other hand, when the Moon scatters an object onto an orbit with a small periapse, the Moon generally gains angular momentum and its semi-major axis increases; this is because the scattered object is typically quickly lost to tidal disruption as it passes close to the Earth, producing a net positive torque on the Moon's orbit. Figure 4b shows the fraction of objects spawned at the Roche limit that merge with the Moon (solid line), that get tidally disrupted after having been scattered inward by the Moon (dotted line) or that get ejected from the system (dashed line).

The fate of a moonlet (accretion or scattering) depends on whether it recoils fast enough from the inner disc to cross the Moon's orbit before its eccentricity gets too high and its pericentre gets inside $2R_\oplus$, leading to its tidal disruption. With time, the mass of the inner disc and the mass of spawned moonlets decrease, resulting in a slower rate of increase of the semi-major axis of objects spawned at the Roche limit, while their eccentricities are excited even faster by the growing Moon. As the 'push' from the inner disc gets weaker with time, an increasing number of objects get scattered inwards and tidally disrupted, causing a net expansion of the Moon's orbit. As the Moon's orbit increases, it becomes more difficult for spawned moonlets to reach its orbit and be accreted. As a result, there is a critical point (at approx. 170 years in figure 4b) beyond which most new objects get scattered inwards and lost as the Moon recedes away (figure 4a, solid line).

After approximately 200 years, a spawned moonlet recoils slowly enough that it can be trapped into inner mean motion resonances with the Moon. The resonant torque due to the disc–moonlet interaction can then be transferred to the Moon's orbit, furthering its expansion. Moonlets trapped in inner resonances with the Moon experience a growth of their eccentricity and are typically eventually lost to tidal disruption. However, once multiple moonlets are trapped together in resonance, their mutual interactions can lead to some being ejected from resonance and collisions with the Moon, finishing its accretion.

**Table 2.** Simulation results. $M$, $f$, $a$ and $e$ are the mass, mass fraction of inner disc material, semi-major axis and eccentricity, respectively, of the largest Moon at a simulation time of 1000 years. $M_2$, $f_2$, $a_2$ and $e_2$ are the same quantities, but for the second largest body in the simulation. Units of mass, distance and angular momentum are the present lunar mass $M_L$, Earth radius $R_\oplus$ and angular momentum of the Earth–Moon system $L_{EM} = 3.5 \times 10^{34}$ kg m$^{-2}$ s$^{-1}$.

| run | $M$ ($M_L$) | $f$ (%) | $a$ ($a_R$) | $e$ | $M_2$ ($M_L$) | $f_2$ (%) | $a_2$ ($a_R$) | $e_2$ |
|---|---|---|---|---|---|---|---|---|
| 1 | 0.294 | 16.3 | 2.98 | 0.070 | 0.202 | 100 | 1.89 | 0.327 |
| 2 | 0.573 | 60 | 2.09 | 0.017 | 0.002 | 100 | 1.29 | 0.261 |
| 3 | 0.531 | 56 | 2.43 | 0.011 | 0.001 | 100 | 1.14 | 0.017 |
| 4 | 0.671 | 30.1 | 2.18 | 0.004 | 0.004 | 100 | 1.37 | 0.254 |
| 5 | 0.699 | 36.6 | 2.10 | 0.002 | 0.002 | 100 | 1.3 | 0.192 |
| 6 | 0.582 | 28.8 | 2.04 | 0.032 | 0.045 | 100 | 1.27 | 0.394 |
| 7 | 0.788 | 20.3 | 2.17 | 0.001 | 0.003 | 100 | 1.30 | 0.144 |
| 8 | 0.631 | 5 | 1.91 | 0.001 | 0.001 | 100 | 1.20 | 0.247 |
| 9 | 0.712 | 12.1 | 2.72 | 0.021 | 0.036 | 100 | 1.27 | 0.013 |
| 10 | 0.356 | 34.3 | 3.04 | 0.095 | 0.260 | 100 | 1.89 | 0.320 |
| 11 | 0.341 | 37.9 | 3.36 | 0.161 | 0.249 | 100 | 2.12 | 0.486 |
| 12 | 0.408 | 100 | 2.03 | 0.180 | 0.243 | 19.4 | 3.32 | 0.219 |
| 13 | 0.567 | 18.2 | 2.67 | 0.028 | 0.183 | 100 | 1.67 | 0.250 |
| 14 | 0.895 | 48.4 | 2.02 | 0.002 | 0.002 | 100 | 1.25 | 0.359 |
| 15 | 0.869 | 48.1 | 2.11 | 0.002 | 0.003 | 100 | 1.32 | 0.444 |
| 16 | 1.018 | 33.3 | 2.10 | 0.001 | 0.002 | 100 | 1.30 | 0.296 |
| 17 | 0.691 | 15.1 | 2.64 | 0.052 | 0.161 | 100 | 1.67 | 0.282 |
| 18 | 0.776 | 17.7 | 2.82 | 0.036 | 0.034 | 100 | 1.76 | 0.504 |
| 19 | 0.396 | 40 | 2.97 | 0.119 | 0.346 | 100 | 1.88 | 0.359 |
| 20 | 0.531 | 56 | 2.68 | 0.001 | 0.226 | 100 | 1.66 | 0.150 |
| 21 | 0.610 | 61.1 | 2.77 | 0.057 | 0.069 | 100 | 1.32 | 0.357 |
| 22 | 0.983 | 51.7 | 2.00 | 0.001 | 0.002 | 100 | 1.24 | 0.387 |
| 23 | 1.047 | 61.2 | 2.01 | 0.001 | 0.002 | 100 | 1.27 | 0.239 |
| 24 | 1.024 | 54.4 | 2.17 | 0.001 | 0.004 | 100 | 1.33 | 0.334 |
| 25 | 0.985 | 36.3 | 2.05 | 0.002 | 0.002 | 100 | 1.28 | 0.422 |
| 26 | 0.946 | 31.7 | 2.07 | 0.001 | 0.002 | 100 | 1.30 | 0.405 |
| 27 | 1.008 | 39.1 | 2.07 | 0.001 | 0.002 | 100 | 1.22 | 0.081 |
| 28 | 0.275 | 53.1 | 2.96 | 0.108 | 0.158 | 100 | 1.87 | 0.409 |
| 29 | 0.53 | 58.6 | 2.26 | 0.016 | 0.003 | 100 | 1.38 | 0.097 |
| 30 | 0.402 | 59.8 | 2.62 | 0.023 | 0.135 | 100 | 1.65 | 0.204 |
| 31 | 0.303 | 40.5 | 3.13 | 0.053 | 0.256 | 100 | 1.89 | 0.384 |

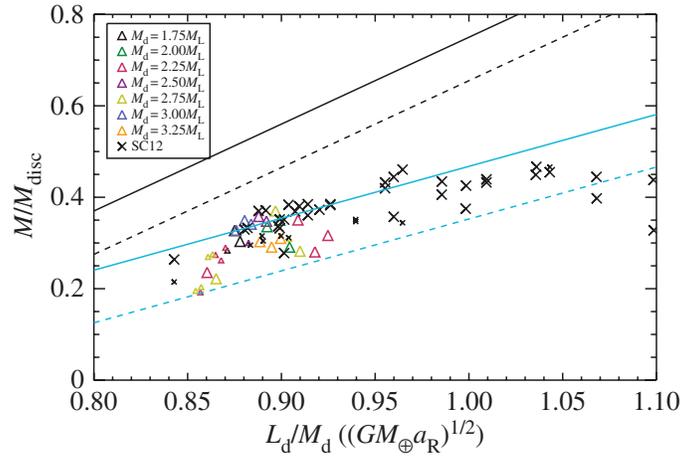

**Figure 5.** Mass of the largest orbiting body at $t = 1000$ years, normalized to the total initial disc mass, against the disc's initial total specific angular momentum. Triangle symbols are the non-canonical disc cases, with colours representing the total initial disc's mass. Black crosses are results from [19]. Black and blue lines are analytical estimates from [11,19], respectively. Solid lines are obtained for an escaping mass of 0, whereas dashed lines assume an escaping mass of 0.05 $M_d$. Non-canonical discs are generally less extended, and thus have lower specific angular momentum than canonical discs.

### (ii) Global results

Table 2 shows the mass, semi-major axis, eccentricity and mass fraction of inner disc material for the Moon and the second largest body at $t = 1000$ years. The average semi-major axis of the Moon is $\langle a \rangle = 2.4\,a_R \pm 0.4$, slightly higher than the average of approximately $2.2\,a_R$ found in [19].

The average mass of the Moon is $\langle M \rangle = 0.66 \pm 0.25 M_L$, with an average mass fraction of inner disc material $\langle f \rangle = 41 \pm 20\%$. The average mass of the Moon is smaller than $\langle M \rangle = 0.81 \pm 0.21 M_L$ obtained with discs in [19]. This is because non-canonical discs have a higher portion of their total mass located in the inner disc. While accretion in the outer disc is fairly efficient, accretion of material from the inner disc is a self-limiting process, because each object spawned at the disc's edge will typically confine it inside the Roche limit via resonant interactions. During this time, the inner disc loses mass on the Earth through its inner edge as it viscously spreads, while no material is provided to the forming Moon. For instance, in run 22, the 'core' formed from the material initially in the outer disc has a mass of $0.47 M_L$, representing 95% of the mass initially present in the outer disc. The mass from the inner disc added to the Moon subsequently totals $0.51 M_L$, which amounts to only 20% of the mass initially in the inner disc.

Figure 5 shows the mass of the Moon at $t = 1000$ years, normalized to the total initial disc mass, against the disc's total initial specific angular momentum. For comparison, we have added the data points for discs simulated in [19] (black crosses). The black and blue lines are analytical estimates assuming that the Moon forms at $1.3\,a_R$ [11] and $2.15\,a_R$ [19], respectively. Solid lines assume an escaping mass $M_{esc}$ of 0, whereas dashed lines assume an escaping mass of 5% of the total initial disc mass. The analytical estimates from [19] are in good agreement with results from the more compact disc cases here and predict a final Moon mass

$$\frac{M_M}{M_d} = 1.14 J_d - 0.67 - 2.3 \frac{M_{esc}}{M_d}. \tag{3.1}$$

## 4. Discussion

### (a) Summary

Using a hybrid numerical lunar disc model developed in [19], we have studied the accretion of the Moon from radially compact protolunar discs similar to those produced by non-canonical impacts, which leave the Earth–Moon system with a large excess in angular momentum that must then be removed by prolonged capture into the so-called evection resonance with the Sun [20,21]. Similar overall accretion dynamics to those seen in less compact discs typical of canonical impacts is seen, with the Moon forming in three consecutive phases over a total period of approximately $10^2$ years [19].

Specifically, we find that, for compact discs with normalized specific angular momenta in the range $0.85 < J_d \leq 0.92$, the fraction of the initial disc mass incorporated into the Moon is between 19% and 37%. This is lower than that predicted using the analytical relation derived from pure $N$-body simulations of [11], which is approximately 45–55%. In the model presented here, the inner disc lifetime is much longer (approx. $10^2$ years) as would be appropriate for a thermally regulated two-phase silicate disc [16,17], and this allows for a prolonged period of interaction between the growing Moon and the inner disc and inner moonlets spawned from the disc. Such interactions generally lead to a larger lunar semi-major axis, which through conservation arguments reduces the mass of the Moon for a given initial disc mass and angular momentum [19].

Our final moon masses here and in [19] are consistent with equation (3.1) with $M_{esc}/M_d \leq 0.05$. Comparing this estimate with the results of successful fast-spinning Earth impacts in [21], we find that very few of the discs are predicted to yield a lunar-mass Moon. Indeed, apart from run 16, forming a lunar-mass object seems to require $M_d \geq 3M_L$, which is somewhat beyond the disc masses obtained by Ćuk & Stewart [21]. For the successful half-Earth impactor cases identified in [20], we find that about half could produce a lunar-mass Moon by equation (3.1) assuming $M_{esc} = 0$.

In the simulations here, the Moon's eccentricity at the end of its accretion is small, of order approximately $10^{-3}$ in cases with $M_M \geq M_L$ (although see the discussion in the next section). If the Moon's orbital expansion due to planetary tides is sufficiently slow as it passes through the resonance (i.e. with tidal parameters for the Earth $(k_2/Q) \leq 5 \times 10^{-3}$ per [21]), such a low eccentricity should ensure initial capture into the evection resonance if the Moon's initial position is well interior to the resonance [22]. However, the simulations performed here produce a Moon whose average semi-major axis by the end of its accretion is approximately $7R_\oplus \pm 1.2$, which is comparable to the position of the evection resonance for an Earth rotating with a period of approximately 2.5 h in the absence of a disc ($6.8R_\oplus$ per [21]). The presence of the disc will shift the location of the evection resonance outwards, but as the disc dissipates the resonance will sweep inwards until reaching this distance. Thus, it appears that the resonance may sweep past the Moon as it is in its final stages of formation and the disc is dissipating. In this case, the relative rate of motion between the Moon and the resonant location may be set by the rate of the disc's dissipation, rather than by the rate of the Moon's orbital expansion due to tides as has been considered previously. This may lead to a substantially reduced likelihood of capture into evection and should be evaluated.

## (b) Model limitations

Our treatment of the Roche interior disc is highly simplified, and in particular it assumes a uniform surface density and that the gas and liquid coevolve. A more physically realistic treatment could affect the accretion efficiency from the disc as well as the accretion time scale (see discussion in [19]).

By modelling the Roche exterior disc by a collection of $N$-body particles, we are implicitly assuming that the outer disc is composed primarily of condensed material. While this appears a good assumption for discs produced by canonical impacts [9,10], non-canonical impacts have a substantially higher impact energy and produce discs that are initially primarily vapour. In the absence of viscous dissipation in this region, cooling and condensation of material orbiting beyond the Roche limit may commence soon after the impact; in this case, a silicate vapour outer disc of surface density $\sigma$ can condense on a time scale $\tau_{cool} \sim \sigma l_v/(2\sigma_{SB} T^4)$, which for $l_v = 2 \times 10^{11}$ erg g$^{-1}$, $\sigma = 10^6$ g cm$^{-2}$ (appropriate, for example, for an outer disc with $0.5M_L$ spread uniformly between $a_R$ and $6R_\oplus$) and $T = 2000$ K is a few years. The simulations here would then approximate the disc's evolution starting a few years post-impact, with a key caveat that there could be important evolution of the disc's properties in the interim period due, for example, to the relaxation of an initially steep radial surface density [30] and/or viscous spreading of the outer disc if and while it maintains a marginally unstable state [17] (see also appendix A).

Non-canonical impacts leave the post-impact Earth with a very short rotation period of approximately 2–3 h [20,21]. Such a rapid rotation implies a highly oblate early Earth, with Earth's $J_2$ proportional to the square of its angular frequency. While the current Earth has $J_2 \sim 10^{-3}$, with a 2–3 h day $J_2$ increases to of order $10^{-1}$. This will lead to more rapid precession of the orbits of objects growing in the disc. The presence of an inner disc will also drive orbital precession in outer growing moons in a time-dependent manner as the disc evolves (e.g. [31]). Neither of these effects has been included in our simulations here, and they could affect the Moon's final orbital properties, for example, if they alter the probability that spawned moonlets are captured into mean motion resonances with the Moon. We have also neglected the effects of tidal dissipation in the Earth and the moonlets. The former acts on time scales of order $10^3$–$10^4$ years (for an Earth Love number $K_{2\oplus} \approx 1$, a tidal dissipation factor $Q_\oplus \sim 10$–$100$ and an orbiting object with a mass $m \sim M_L$) and may not be significant over accretion time scales. The latter, however, can lead to damping of the eccentricities of the orbiting bodies over time scales comparable to accretion time scales. Both effects could affect resonant capture likelihood as well as the Moon's orbital state at the end of its assembly and will be included in further modelling.

Acknowledgements. We thank Alex Halliday and David Stevenson for their invitation to the 'Origin of the Moon' meeting. We thank Marc Buie and David Kaufmann for valuable discussions on the Earth's oblateness.
Funding statement. Support from NASA is gratefully acknowledged.

## Appendix A

A self-gravitating Keplerian disc of surface density $\sigma$ becomes susceptible to instability for $(c/c_{crit}) < 1$, where $c$ is the relevant disc velocity (i.e. the sound speed for a vapour disc or the dispersion velocity for a particulate disc), $c_{crit} \equiv \pi G \sigma / \Omega$ is the critical velocity for instability and $\Omega$ is the orbital frequency (e.g. [16]). For $c < c_{crit}$, instabilities grow on a time scale $\tau_{inst} = \Omega^{-1}[(c_{crit}/c)^2 - 1]^{-1/2}$. Interior to the Roche limit, clumps in the disc produced by instabilities are continually tidally disrupted, producing a viscosity and associated heating in this inner disc region [16].

Exterior to the Roche limit, whether clumps formed by local instabilities survive or are disrupted depends on how fast they form compared with the rate at which they are sheared apart by the disc's differential Kepler motion. If $\tau_{inst}$ is shorter than the orbital time scale (where $\tau_{orb} \sim (\frac{2}{3})\Omega^{-1}$), clumps formed by instabilities exterior to the Roche limit are stable and the disc fragments [16]. This is the case for unstable discs having $(c/c_{crit}) < 0.6$. However, if a disc is only marginally unstable, specifically with $0.6 \ll (c/c_{crit}) < 1$, then $\tau_{inst} \gg \tau_{orb}$ and clumps formed by instabilities exterior to the Roche limit are sheared apart by differential rotation, producing a viscosity in this region as well [17].

Thompson and Stevenson [17] consider a protolunar disc in which gas and melt phases remain well mixed and the two-phase sound speed is regulated to be just below $c_{crit}$, so that the disc maintains a marginally unstable, 'metastable' state. In this case $(c/c_{crit}) \sim 1$, and clumps in the Roche exterior disc form slowly enough that they can be disrupted by shear, producing a viscosity in the outer disc. As the disc spreads, $\sigma$ decreases, which causes the disc to become more unstable (i.e. reducing its $(c/c_{crit})$ ratio) in order to maintain the assumed local balance between viscous dissipation and radiative cooling by the photosphere. Eventually as $\tau_{inst} \sim \tau_{orb}$, the disc fragments. The distance at which fragmentation occurs then migrates inwards with time until it reaches the Roche limit [17].

Ward [18] considers a stratified protolunar disc in which a mid-plane magma layer coexists with an extended two-phase, vapour-rich atmosphere. In this case, the magma layer is unstable rather than metastable, so that instabilities outside the Roche limit form rapidly, rendering them stable against disruption by shear.

The accretion models here and in [19] assume that there is no instability-induced viscosity in the Roche exterior disc. This is consistent with the stratified disc model of [18], and the later evolution of the disc in the model of [17].